# Differential Equation Approach for One- and Two-Dimensional Lattice Green's Function

**J. H. Asad**

Dep. of Phys. University of Jordan- amman 11942.

E-mail: jhasad1@yahoo.com.

**Abstract**

A first order differential equation of Green's function, at the origin G (0), for the one-dimensional lattice is derived by simple recurrence relation. Green's function at site (m) is then calculated in terms of G (0). A simple recurrence relation connecting the lattice Green's function at the site (m, n) and the first derivative of the lattice Green's function at the site (m±1,n) is presented for the two- dimensional lattice, a differential equation of second order in G (0,0) is obtained. By making use of the latter recurrence relation, lattice Green's function at an arbitrary site is obtained in closed form. Finally, the phase shift and scattering cross section are evaluated analytically and numerically for one- and two impurities.

***Key Words***: *Green's Function, One- and two-dimensional lattices,* impurity.

## I- INTRODUCTION

The Lattice green's Function (LGF) is a basic function in the study of the solid state physics and condensed matter physics. It appears especially when impure solids are studied[1-5]. Nowadays, Green's function (GF) becomes one of the most important concepts in many branches of physics, because many quantities of interest can be expressed in terms of LGF, and in the following are some examples:

- i- Statistical Model of Ferromagnetism such as Ising Model[6].
- ii- Random Walk Theory[7].
- iii- Diffusion[8].
- iv- Heisenberg Model[9].
- v- Band structure[10,11].
- vi- Resistance Calculation for an Infinite Networks of Identical Resistors[12-16].

The LGF for several structure lattices has been widely studied by many authors[17-21]. In a recent work GF, density of states (DOS), phase shift, and scattering cross section have been evaluated numerically and analytically for different cases, and below are some examples:

1- General Glasser Case lattice[3].
2- Face Centered Cubic (FCC) Lattice[4]
3- Glasser Case Lattice[22].
4- Body centered Cubic (BCC) Lattice[23].

In this paper; GF at the origin is expressed as a first order differential equation for one- dimensional lattice, which enables us to calculate GF at an arbitrary site using the so- called recurrence formulae. On the other hand GF for the two- dimensional lattice at an arbitrary site is obtained in a closed form, which contains a sum of the complete elliptic integrals of the first and second kind.

Finally, the phase shift, scattering cross section and DOS are evaluated numerically and analytically for one- and two- impurities.

**II- Green's Function for One Dimensional Lattice**

The LGF for one-dimensional lattice is defined by[1]

$$G(m,t) = \frac{1}{\pi^2}\int_0^{\pi} \frac{\mathrm{Cosmx}}{t - \mathrm{Cosx}}\, dx\,, \tag{1}$$

Where m is integer and t is a parameter.

By executing the integration with respect to x in Eq. (1), we obtained the following recurrence relation:

$$G'(m+1) - G'(m-1) = 2mG(m), \tag{2}$$

Where $G'(m)$ expresses the first derivative of G(m) with respect to t.

Taking the derivatives of Eq. (2) with respect to t, we obtained recurrence relations involving higher derivatives of GF.

For m=1 and 2, we obtained:

$$G'(2) - G'(0) = 2G(1), \tag{3}$$

$$G'(3) - G'(1) = 4G(2), \tag{4}$$

Noting the following recurrence relation[20,24] :

$$G(m+1) = 2tG(m) - G(m-1) - 2\delta_{m0}, \tag{5}$$

for m=0 we find the well-known relation:

$$G(1) = tG(0) - 1. \tag{6}$$

For m=1 and 2, we have[24,25]

$$G(2) = 2tG(1) - G(0), \tag{7}$$

$$G(3) = 2tG(2) - G(1), \tag{8}$$

Taking the derivative of both sides of Eq. (8) with respect to t, and make use of Eqs. (3), (4) and (6), we obtained the following expressions:

$$G(2) = (2t^2 - 1)G(0) - 2t, \tag{9}$$

$$G(3) = (4t^3 - 3t)G(0) + (1 - 4t^2), \tag{10}$$

Similarly, derive both sides of Eq. (9) with respect to t, and using Eqs. (3) and (6), we obtained the following differential equation for G(0):

$$G'(0) + \frac{t}{t^2 - 1} G(0) = 0, \tag{11}$$

Integrating the above equation one get

$$G(0,t) = \frac{1}{\sqrt{t^2 - 1}}. \tag{12}$$

So, the values of the one-dimensional LGF at an arbitrary site are known. The diagonal LGF inside and outside the band for one-dimensional lattice is

$$G^0(\mathrm{L},\mathrm{L},\mathrm{E}) = \begin{cases} \dfrac{i}{(1-\mathrm{E}^2)^{1/2}}, & |\mathrm{E}| < 1 \\ \dfrac{1}{(\mathrm{E}^2-1)^{1/2}}, & |\mathrm{E}| > 1 \end{cases} \tag{13}$$

Now, let us consider the tight- binding Hamiltonian (TBH) with the following two cases:

**a- single impurity atom**

We consider here a TBH whose perfect periodicity is destroyed due to the presence of the point defect at the L site. This situation can be thought of physically as arising by substituting the host atom at the L-site by a foreign atom[26] having a level lying $\varepsilon'$ higher than the common level of the host atoms (L), also, taking into account the interaction between the point defect with the first nearest neighboring host atoms.

After some mathematical manipulations the single impurity GF can be written as[1,2]:

$$G(L,E) = \begin{cases} \dfrac{1}{(1+2\beta)(E^2-1)^{1/2} - (\varepsilon' + 2\beta E)}, & |E| > 1 \\[2ex] \dfrac{-(\varepsilon' + 2\beta E) + i(1+2\beta)(1-E^2)^{1/2}}{(1+2\beta)(1-E^2) + (\varepsilon' + 2\beta E)^2}, & |E| < 1 \end{cases} \qquad (14)$$

Therefore, the local DOS at site L has the form:

$$DOS(E) = \frac{(1+2\beta)(1-E^2)^{1/2}}{\pi\,[(1+2\beta)(1-E^2) + (\varepsilon' + 2\beta E)^2]}, \qquad (15)$$

The S-wave phase shift, $\delta_0$, is defined as[1]:

$$\tan\delta_o = \frac{\operatorname{Im} G^0(E)}{\dfrac{1+2\beta}{\varepsilon' + 2\beta E} - \operatorname{Re} G^0(E)}, \qquad (16)$$

Here, ReG $^0$ (E) refers to the real part of GF inside the band, $\varepsilon'$ is a constant depending on the strength of the impurity potential, and β is the interaction parameter (hopping integral). After some mathematical manipulations, we obtained the S - wave shift as:

$$\tan\delta_0 = \frac{\varepsilon' + 2\beta E}{(1+2\beta)(1-E^2)^{1/2}}, \tag{17}$$

The cross- section, σ, is defined as[1]:

$$\sigma = \frac{4\pi}{P^2} \frac{(\varepsilon' + 2\beta E)^2 [\operatorname{Im} G^0(E)]^2}{\left[(1+2\beta) - (\varepsilon' + 2\beta E)\operatorname{Re} G^0(E)\right]^2 + (\varepsilon' + 2\beta E)^2 [\operatorname{Im} G^0(E)]^2}, \tag{18}$$

Here, $P$ refers to the electron momentum.

Therefore, the cross- section becomes

$$\sigma = \frac{4\pi}{p^2} \frac{(\varepsilon' + 2\beta E)^2}{(1+2\beta)^2 (1-E^2) + (\varepsilon' + 2\beta E)^2}. \tag{19}$$

**b- two impurity atoms**

In the following we consider the case where two substitutional impurities are introduced at two different sites (l and m) of the lattice. The GF at arbitrary site (n,n) is[2]

$$G(\vec{n},\vec{n}) = G_0(\vec{n},\vec{n}) + \frac{\varepsilon'_l G_0^2(\vec{l},\vec{l})(1-\varepsilon'_m G_0(\vec{m},\vec{m})) + \varepsilon'_m G_0^2(\vec{m},\vec{m})(1-\varepsilon'_l G_0(\vec{l},\vec{l})) + 2\varepsilon'_m \varepsilon'_l G_0^3(\vec{l},\vec{m})}{(1-\varepsilon'_l G(\vec{l},\vec{l}))(1-\varepsilon'_m G(\vec{m},\vec{m})) - \varepsilon'_m \varepsilon'_l G_0^2(\vec{l},\vec{m})}, \tag{20}$$

After some mathematical manipulations the GF can be written as:

$$G(\vec{n},\vec{n}) = \frac{E^2 - 1 + \varepsilon'_m \varepsilon'_l [(E - \sqrt{E^2-1})^2 (2E - 1 - 2\sqrt{E^2-1}) - 1]}{[(E^2-1)^{\frac{3}{2}} (1 - 2\varepsilon'_l \varepsilon'_m) - (\varepsilon'_l + \varepsilon'_m - 2E\varepsilon'_m \varepsilon'_l)(E^2-1)]}. \tag{21}$$

The S-wave phase shift, $\delta_0$, is defined as:

$$\tan\delta_0 = \frac{\varepsilon_l' \operatorname{Im} G_0(l,l)(1-\varepsilon_m' \operatorname{Re} G_0(m,m)) + \varepsilon_m' \operatorname{Im} G_0(m,m)(1-\varepsilon_l' \operatorname{Re} G_0(l,l)) + 2\varepsilon_m'\varepsilon_l' \operatorname{Re} G_0(l,m) \operatorname{Im} G_0(l,m)}{(1-\varepsilon_l' \operatorname{Re} G_0(l,l))(1-\varepsilon_m' \operatorname{Re} G_0(m,m)) - \varepsilon_m'\varepsilon_l'[\operatorname{Im} G_0(l,l) \operatorname{Im} G_0(m,m) + (\operatorname{Re} G_0(l,m))^2 - (\operatorname{Im} G_0(l,m))^2]}, \quad (22)$$

Therefore, phase shift, $\delta_0$, is

$$\tan\delta_0 = \frac{\varepsilon_l' + \varepsilon_m' - 2\varepsilon_l'\varepsilon_m' E}{(1-2\varepsilon_l'\varepsilon_m')(1-E^2)^{1/2}}, \quad (23)$$

Then, the cross section can be written as:

$$\sigma = \frac{4\pi}{p^2} \frac{1}{\left[\frac{(1-\varepsilon_l' \operatorname{Re} G_0(l,l))(1-\varepsilon_m' \operatorname{Re} G_0(m,m)) - \varepsilon_m'\varepsilon_l'(\operatorname{Im} G_0(l,l) \operatorname{Im} G_0(m,m) + (\operatorname{Re} G_0(l,m))^2 - (\operatorname{Im} G_0(l,m))^2)}{\varepsilon_l' \operatorname{Im} G_0(l,l)(1-\varepsilon_m' \operatorname{Re} G_0(m,m)) + \varepsilon_m' \operatorname{Im} G_0(m,m)(1-\varepsilon_l' \operatorname{Re} G_0(l,l)) + 2\varepsilon_l'\varepsilon_m' \operatorname{Re} G_0(l,m) \operatorname{Im} G_0(l,m)}\right]^2 + 1}, \quad (24)$$

Therefore, the cross- section becomes

$$\sigma = \frac{4\pi^2}{p^2} \frac{(\varepsilon_l' + \varepsilon_m' - 2\varepsilon_l'\varepsilon_m' E)^2}{(1-2\varepsilon_l'\varepsilon_m')^2(1-E^2) + (\varepsilon_l' + \varepsilon_m' - 2\varepsilon_l'\varepsilon_m' E)^2}. \quad (25)$$

**III- Green's Function for Two Dimensional Lattice**

The LGF for two-dimensional lattice is defined by[1]

$$G(m,n,t) = \frac{1}{\pi^2}\int_0^\pi \int_0^\pi \frac{\operatorname{Cos} mx \operatorname{Cos} ny}{t-(\operatorname{Cos} x + \operatorname{Cos} y)}\, dx\, dy, \quad (26)$$

Where (m, n) are integers and t is a parameter.

By executing a partial integration with respect to x in Eq. (26), we obtained the following recurrence relation[20]:

$$G'(m+1,n) - G'(m-1,n) = 2mG(m,n), \tag{27}$$

where $G'(m,n)$ expresses the first derivative of G(m,n) with respect to t. Taking derivatives of Eq. (27) with respect to t, we obtained recurrence relations involving higher derivatives of the GF.

Putting (m,n)=(1,0), (1,1), and (2,0) in Eq. (27), respectively we obtained the following relations:

$$G'(2,0) - G'(0,0) = 2G(1,0), \tag{28}$$

$$G'(2,1) - G'(1,0) = 2G(1,1), \tag{29}$$

$$G'(3,0) - G'(1,0) = 4G(2,0). \tag{30}$$

For m=0 we obtain[3-5]

$$2tG(0,n) - 2\delta_{0n} - 2G(1,n) - G(0,n+1) - G(0,n-1) = 0 \tag{31}$$

Insert n=0 in Eq. (31) we find the well-known relation

$$G(1,0) = \frac{1}{2}[tG(0,0) - 1], \tag{32}$$

for m≠0 we have

$$G(m+1,n) - 2tG(m,n) + G(m-1,n) + G(m,n+1) + G(m,n-1) = 0, \tag{33}$$

Substituting (m,n)=(1,0), (1,1), and (2,0) in Eq. (33), respectively we obtained the following relations:

$$G(1,1) = tG(1,0) - \frac{1}{2}G(0,0) - \frac{1}{2}G(2,0), \tag{34}$$

$$G(2,1) = (t^2 - 1)G(1,0) - \frac{t}{2}G(0,0) - \frac{t}{2}G(2,0), \tag{35}$$

$$G(3,0) = (\frac{3}{2}t - t^3)G(0,0) + 3tG(2,0) - (\frac{1-2t^2}{2}), \tag{36}$$

Now, by taking the derivative of both sides of Eq. (36) with respect to t, and using Eqs. (28), (29) and (30), we obtained the following expressions:

$$G(2,0) = (4t - t^3)G'(0,0) + G(0,0) - t, \tag{37}$$

$$G(1,1) = (\frac{t^2}{2} - 1)G(0,0) - \frac{t}{2}(4 - t^2)G'(0,0), \tag{38}$$

$$G(2,1) = \frac{t}{2}(t^3 - 3)G(0,0) - \frac{t^2}{2}(4 - t^2)G'(0,0) - \frac{1}{2}, \tag{39}$$

$$G(3,0) = \frac{t}{2}(9 - 2t^2)G(0,0) + 3t^2(4 - t^2)G'(0,0) - (\frac{1 + 4t^2}{2}), \tag{40}$$

Again, taking the derivative of both side of Eq. (37) with respect to t, and using Eqs. (28) and (32), we obtained the following differential equation for G(0,0):

$$t(4 - t^2)G''(0,0) + (4 - 3t^2)G'(0,0) - tG(0,0) = 0. \tag{41}$$

Where $G^{//}(0,0)$ is the second derivative of G(0,0).

By using the following transformations G(0,0) = Y(x)/t and $x=4/t^2$ we obtain the following differential equation[27-29]:

$$x(1 - x)\frac{d^2Y(x)}{dx^2} + (1 - 2x)\frac{dY(x)}{dx} - \frac{1}{4}Y(x) = 0, \tag{42}$$

This is called the hypergeometric differential equation (Gauss's differential equation). So, the solution is[29]

$Y(x)={}_1F_2(1/2,1/2;1;x) = (2/\pi)\ K(2/t)$

Then,

$$G(0,0,t) = \frac{2}{\pi t}K(\frac{2}{t}) \tag{43}$$

By using Eq. (43) we can express $G^{/}(0,0)$ and $G^{//}(0,0)$ in terms of the complete elliptic integrals of the first and second kind.

$$G'(0,0,t) = \frac{2}{\pi}\frac{E(\frac{2}{t})}{4-t^2}, \qquad (44)$$

$$G''(0,0,t) = \frac{2}{\pi t(t^2-4)}[E(\frac{2}{t})[3t^2-4] - K(\frac{2}{t})]. \qquad (45)$$

K(2/t) and E(2/t) are the complete elliptic integrals of the first and second kind, respectively. So that, the two-dimensional LGF at an arbitrary site is obtained in closed form, which contains a sum of the complete elliptic integrals of the first and second kind.

The diagonal GF outside and inside the band becomes:

$$G^0(\mathrm{L,L,E}) = \begin{cases} \dfrac{2}{\pi\,\mathrm{E}}\mathrm{K(2/E)} & ; \ |\mathrm{E}| > 2 \\ \dfrac{1}{\pi}\left[\mathrm{K}(E/2) + \mathrm{iK}(\sqrt{1-E^2/4})\right] & ; \ |\mathrm{E}| < 2 \end{cases} \qquad (46)$$

Again, let us consider the following two cases:

**a- single impurity atom**

For a single substitutional impurity at site L the defect GF for the square lattice is[1,2]:

$$G^0(\mathrm{L,L,E}) = \begin{cases} \dfrac{\mathrm{K(E/2)}}{\frac{\pi \mathrm{E}}{2}(1+2\beta) - (\varepsilon' + 2\beta E)K(E/2)} & ; \ |\mathrm{E}| > 2 \\ \dfrac{[\pi(1+2\beta) - (\varepsilon'+2\beta E)K(E/2)]K(E/2) + [i\pi(1+2\beta) - (\varepsilon'+2\beta E)K(\sqrt{1-E^2/4})]K(\sqrt{1-E^2/4})}{[\pi(1+2\beta) - (\varepsilon'+2\beta E)K(E/2)]^2 + (\varepsilon'+2\beta E)^2K^2(\sqrt{1-E^2/4})} & ; \ |\mathrm{E}| < 2 \end{cases} \qquad (47)$$

Therefore, the DOS is

$$\mathrm{DOS(E)} = \begin{cases} \dfrac{(1+2\beta)K(\sqrt{1-E^2/4})}{[\pi(1+2\beta) - (\varepsilon'+2\beta E)K(E/2)]^2 + (\varepsilon'+2\beta E)^2K^2(\sqrt{1-E^2/4})} & ; \ |\mathrm{E}| < 2 \end{cases} \qquad (48)$$

The S-wave phase shift, $\delta_0$, is defined as :

$$\tan\delta_o = \frac{\operatorname{Im} G^0(E)}{\frac{1+2\beta}{\varepsilon' + 2\beta E} - \operatorname{Re} G^0(E)}, \tag{49}$$

Therefore, phase shift, $\delta_0$, is

$$\tan\delta_o = \frac{K(\sqrt{1-E^2/4})}{\pi \frac{1+2\beta}{\varepsilon' + 2\beta E} - K(E/2)}, \tag{50}$$

The cross- section, $\sigma$, is defined as:

$$\sigma = \frac{4\pi}{P^2} \frac{(\varepsilon' + 2\beta E)^2 [\operatorname{Im} G^0(E)]^2}{\left[(1+2\beta) - (\varepsilon' + 2\beta E)\operatorname{Re} G^0(E)\right]^2 + (\varepsilon' + 2\beta E)^2 [\operatorname{Im} G^0(E)]^2}, \tag{51}$$

Therefore, the cross- section becomes:

$$\sigma = \frac{4\pi}{P^2} \frac{(\varepsilon' + 2\beta E)^2 [K(\sqrt{1-E^2/4})]^2}{[\pi(1+2\beta) - (\varepsilon' + 2\beta E)K(E/2)]^2 + (\varepsilon' + 2\beta E)^2 [K(\sqrt{1-E^2/4})]^2}, \tag{52}$$

**b- two impurity atoms**

We introduced here two substitutional impurities at two different sites of the lattice, $l$, and m.

The GF at site (n,n) is:

$$G(\vec{n},\vec{n}) = G_0(\vec{n},\vec{n}) + \frac{\varepsilon'_l G_0^2(\vec{l},\vec{l})(1-\varepsilon'_m G_0(\vec{m},\vec{m})) + \varepsilon'_m G_0^2(\vec{m},\vec{m})(1-\varepsilon'_l G_0(\vec{l},\vec{l})) + 2\varepsilon'_m \varepsilon'_l G_0^3(\vec{l},\vec{m})}{(1-\varepsilon'_l G(\vec{l},\vec{l}))(1-\varepsilon'_m G(\vec{m},\vec{m})) - \varepsilon'_m \varepsilon'_l G_0^2(\vec{l},\vec{m})}, \tag{53}$$

After some mathematical manipulations the GF can be written as:

$$G(\vec{n},\vec{n}) = \frac{2K(2/E)[1-\frac{4\varepsilon'_l\varepsilon'_m}{\pi^2E^2}K^2(2/E)] + \varepsilon'_m\varepsilon'_l[\frac{K(2/E)}{\pi}-\frac{1}{2}]^2[2(E-1)K(2/E)-\pi E]}{(1-2\varepsilon'_l K(2/E)(1-\frac{2\varepsilon'_m}{\pi E}K(2/E)) - \frac{\pi E\varepsilon'_m\varepsilon'_l}{4}[\frac{2}{\pi}K(2/E)-1]^2}, \tag{54}$$

The S-wave phase shift, $\delta_0$ ,is defined as:

$$\tan\delta_0 = \frac{\varepsilon'_l \operatorname{Im}G_0(\vec{l},\vec{l})(1-\varepsilon'_m \operatorname{Re}G_0(\vec{m},\vec{m})) + \varepsilon'_m \operatorname{Im}G_0(\vec{m},\vec{m})(1-\varepsilon'_l \operatorname{Re}G_0(\vec{l},\vec{l})) + 2\varepsilon'_m\varepsilon'_l \operatorname{Re}G_0(\vec{l},\vec{m})\operatorname{Im}G_0(\vec{l},\vec{m})}{(1-\varepsilon'_l \operatorname{Re}G_0(\vec{l},\vec{l}))(1-\varepsilon'_m \operatorname{Re}G_0(\vec{m},\vec{m})) - \varepsilon'_m\varepsilon'_l[\operatorname{Im}G_0(\vec{l},\vec{l})\operatorname{Im}G_0(\vec{m},\vec{m}) + (\operatorname{Re}G_0(\vec{l},\vec{m}))^2 - (\operatorname{Im}G_0(\vec{l},\vec{m}))^2]}, \tag{55}$$

Therefore, the phase shift, $\delta_0$, is

$$\tan\delta_0 = \frac{[\varepsilon'_l+\varepsilon'_m-\frac{\varepsilon'_l\varepsilon'_m E}{2}-\frac{2\varepsilon'_l\varepsilon'_m}{\pi}(1-E^2/4)K(E/2)]K(\sqrt{1-E^2/4})}{\pi(1-\frac{\varepsilon'_l\varepsilon'_m}{4})-[\varepsilon'_l+\varepsilon'_m-\frac{\varepsilon'_l\varepsilon'_m E}{2}-\varepsilon'_l\varepsilon'_m(1-E^2/4)\frac{K(E/2)}{\pi}]K(E/2)-\frac{\varepsilon_l\varepsilon_m}{\pi}(1-E^2/4)K(\sqrt{1-E^2/4})}, \tag{56}$$

Then, the cross section can be written as

$$\sigma = \frac{4\pi}{p^2}\frac{1}{[\frac{(1-\varepsilon'_l \operatorname{Re}G_0(\vec{l},\vec{l}))(1-\varepsilon'_m \operatorname{Re}G_0(\vec{m},\vec{m})) - \varepsilon'_m\varepsilon'_l(\operatorname{Im}G_0(\vec{l},\vec{l})\operatorname{Im}G_0(\vec{m},\vec{m}) + (\operatorname{Re}G_0(\vec{l},\vec{m}))^2 - (\operatorname{Im}G_0(\vec{l},\vec{m}))^2)}{\varepsilon'_l \operatorname{Im}G_0(\vec{l},\vec{l})(1-\varepsilon'_m \operatorname{Re}G_0(\vec{m},\vec{m})) + \varepsilon'_m \operatorname{Im}G_0(\vec{m},\vec{m})(1-\varepsilon'_l \operatorname{Re}G_0(\vec{l},\vec{l})) + 2\varepsilon'_l\varepsilon'_m \operatorname{Re}G_0(\vec{l},\vec{m})\operatorname{Im}G_0(\vec{l},\vec{m})}]^2+1}, \tag{57}$$

Therefore, the cross- section becomes

$$\sigma = \frac{4\pi}{p^2}\frac{1}{\left(\frac{\pi(1-\frac{\varepsilon'_l\varepsilon'_m}{4})-[\varepsilon'_l+\varepsilon'_m-\frac{\varepsilon'_l\varepsilon'_m E}{2}-\varepsilon'_l\varepsilon'_m(1-E^2/4)\frac{K(E/2)}{\pi}]K(E/2)-\frac{\varepsilon_l\varepsilon_m}{\pi}(1-E^2/4)K(\sqrt{1-E^2/4})}{[\varepsilon'_l+\varepsilon'_m-\frac{\varepsilon'_l\varepsilon'_m E}{2}-\frac{2\varepsilon'_l\varepsilon'_m}{\pi}(1-E^2/4)K(E/2)]K(\sqrt{1-E^2/4})}\right)^2+1}. \tag{58}$$

Finally, as mentioned in the introduction the LGF has many applications [6-11]. The method presented in this paper has an interesting application, where it can be used in calculating the resistance of an infinite network consisting of identical resistors. Where one can express the resistance between the origin and any lattice site in terms of the LGF at the origin and its derivatives[12,15,16] .

## VI- Results and Discussion

First of all, we conclude that using this alternative method one can rewrite the non-diagonal LGF at any lattice site in terms of the LGF at the origin and its derivatives. These values are expressed in turns in terms of the elliptic integrals of the first, and second kind.

The results of the LGF for the one-dimensional lattice are shown in Figs. (1-7) and those of the two-dimensional lattice (square) are shown in Figs. (8-13). Figure 1 shows the DOS for the one-dimensional lattice with single impurity for $\beta = \varepsilon'$, with different potential strengths $\varepsilon'$ (-0.7,-0.3,0.0,0.3, and 0.7), for $\beta = \varepsilon' = -0.7$ we have a discontinuity in the curve as shown in Fig.1.

The phase shift, $\delta_0$, is defined as the shift in the phase of the wave function due to the presence of the impurity potential. Figure 2 displays, $\delta_0$, for the one-dimensional lattice with single impurity for $\beta = \varepsilon'$ with different potential strengths $\varepsilon'$(-0.7,-0.3,0.0,0.3, and 0.7). For $\beta = \varepsilon' = 0.0$, $\delta_0$, vanishes as potential is turned off (perfect lattice). The phase shift vanishes for all potentials values as E goes to –0.5 and is separated into two regions about E= -0.5. Figure 3 shows the phase shift, $\delta_0$, in three dimensions for the one-dimensional lattice with single impurity for different potential strengths $\varepsilon'$ varying between -1 and 1 (arbitrary units).

The cross section, $\sigma$, is defined as the area an impurity atom presents to the incident electron. Figure 4 shows The cross section, $\sigma$, for the one-dimensional lattice with single impurity for $\beta = \varepsilon'$ with different potential strengths $\varepsilon'$(-0.7,-0.3,0.0,0.3, and 0.7). The peak value is a constant for all potential strengths. The cross section can be related to some physical quantities such as the conductivity in metals.

Figure 5 shows the phase shift, $\delta_0$, for the one-dimensional lattice with two identical impurities for different potential strengths $\varepsilon'$ (-0.7, -0.3,0.0,0.3, and 0.7). The curves are mirror images of each other. While, Fig. 6 shows the phase shift, $\delta_0$, in three dimensions for the one-dimensional lattice with two identical impurities for different potential strengths $\varepsilon'$ varying between -1 and 1(arbitrary units).

Figure 7 displays the cross section, $\sigma$, for the one-dimensional lattice with two identical impurities for different potential strengths $\varepsilon'$ (-0.7, -0.3,0.0,0.3, and 0.7). The curves are mirror images of each other. For $\varepsilon' = \pm 0.7$, $\sigma$, has a constant value.

Figure 8 shows the DOS for the square lattice with single impurity for $\beta = \varepsilon'$ with different potential strengths $\varepsilon'$ (-0.7,-0.3,0.0,0.3, and 0.7). For $\varepsilon' = -0.7$ DOS is vanished, the peak value increases as $\varepsilon'$ decreases.

Figure 9 displays the phase shift, $\delta_0$, for the square lattice with single impurity for $\beta = \varepsilon'$ with different potential strengths $\varepsilon'$ (-0.7, -0.3,0.0,0.3, and 0.7). For $\varepsilon' = 0.0$, $\delta_0$ vanishes as the potential is turned off (perfect lattice). We have a discontinuity occurring in the curve as shown in Fig. 9. Figure 10 shows $\delta_0$ in three dimensions for the square lattice with single impurity for different potential strengths $\varepsilon'$ varying between -1 and 1 (arbitrary units), whereas the second axis is energy scale varying between –2 and 2 as indicated in the formalism.

Figure 11 shows The cross section, $\sigma$, for the square lattice with single impurity for $\beta = \varepsilon'$ with different potential strength $\varepsilon'$. Figure12 displays the phase shift, $\delta_0$, for the square lattice with two identical impurities for different potential strengths $\varepsilon'$ (-0.7,-0.3,0.0,0.3, and 0.7). The phase shift is always positive for all negative potentials and vice versa.

Finally, Fig.13 displays the cross section, σ, for the square lattice with two identical impurities for different potential strengths $\varepsilon'$ (-0.7,-0.3,0.0,0.3, and 0.7).The peak value varies with the potential strength and reaches it's a maximum value at $\varepsilon'=0.7$; the peak value increases in range between $0<\varepsilon'<1$ as $\varepsilon'$ increases and decreases otherwise.

## ACKNOWLEDGEMENTS

The author wishes to thank Proff. J. M. Khalifeh, Dr. R. S. Hijjawi, and Dr. A. J. Sakaji for helpful discussion and graphs manipulating.

## Figure Captions

Fig. 1: The density of states (DOS) for the one-dimensional lattice with single impurity for $\beta = \varepsilon'$ with different potential strengths $\varepsilon'$ (-0.7,-0.3,0.0,0.3, and 0.7).

Fig. 2: The phase shift, $\delta_0$, for the one-dimensional lattice with single impurity for $\beta = \varepsilon'$ with different potential strengths $\varepsilon'$(-0.7,-0.3,0.0,0.3, and 0.7).

Fig. 3: The phase shift, $\delta_0$, in three dimensions for the one-dimensional lattice with single impurity for $\beta = \varepsilon'$ with different potential strengths $\varepsilon'$ varying between -1 and 1(arbitrary units ).

Fig. 4: The cross section, $\sigma$, for the one-dimensional lattice with single impurity for $\beta = \varepsilon'$ with different potential strengths $\varepsilon'$(-0.7,-0.3,0.0,0.3, and 0.7).

Fig. 5: The phase shift, $\delta_0$, for the one-dimensional lattice with two identical impurities for different potential strengths $\varepsilon'$(-0.7, -0.3,0.0,0.3, and 0.7).

Fig. 6: The phase shift, $\delta_0$ , in three dimensions for the one-dimensional lattice with two identical impurities for different potential strengths $\varepsilon'$ varying between -1 and 1(arbitrary units ).

Fig. 7: The cross section, $\sigma$, for the one-dimensional lattice with two identical impurities for different potential strengths $\varepsilon'$(-0.7,-0.3,0.0,0.3, and 0.7).

Fig. 8: The density of states (DOS) for the square lattice with single impurity for $\beta = \varepsilon'$ with different potential strengths $\varepsilon'$ (-0.7,-0.3,0.0,0.3, and 0.7).

Fig. 9: The phase shift, $\delta_0$, for the square lattice with single impurity for $\beta = \varepsilon'$ with different potential strengths $\varepsilon'$(-0.7,-0.3,0.0,0.3, and 0.7).

Fig.10: The phase shift, $\delta_0$ , in three dimensions for the square lattice with single impurity for $\beta = \varepsilon'$ with different potential strengths $\varepsilon'$ varying between -1 and 1(arbitrary units ).

Fig.11: The cross section, $\sigma$, for the square lattice with single impurity for $\beta = \varepsilon'$ with different potential strengths $\varepsilon'$(-0.7,-0.3,0.0,0.3, and 0.7).

Fig.12: The phase shift, $\delta_0$, for the square lattice with two identical impurities for different potential strengths $\varepsilon'$(-0.7, -0.3,0.0,0.3, and 0.7).

Fig. 13: The cross section, $\sigma$, for the square lattice with two identical impurities for different potential strengths $\varepsilon'$(-0.7, -0.3,0.0,0.3, and 0.7).

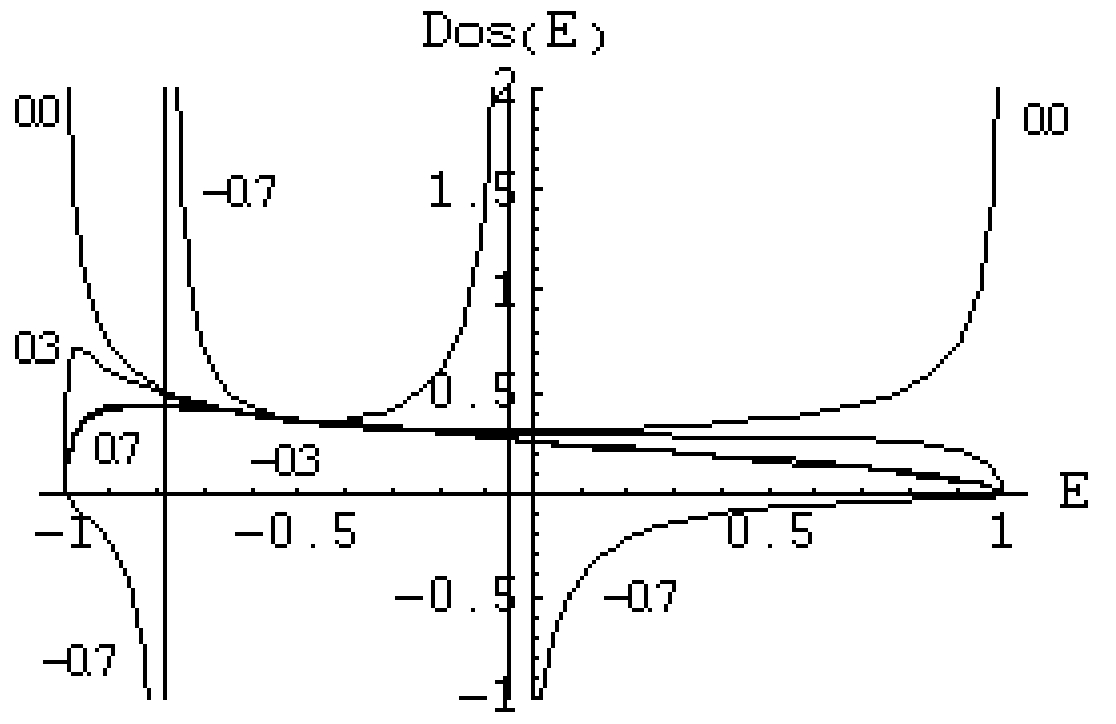

Fig.1

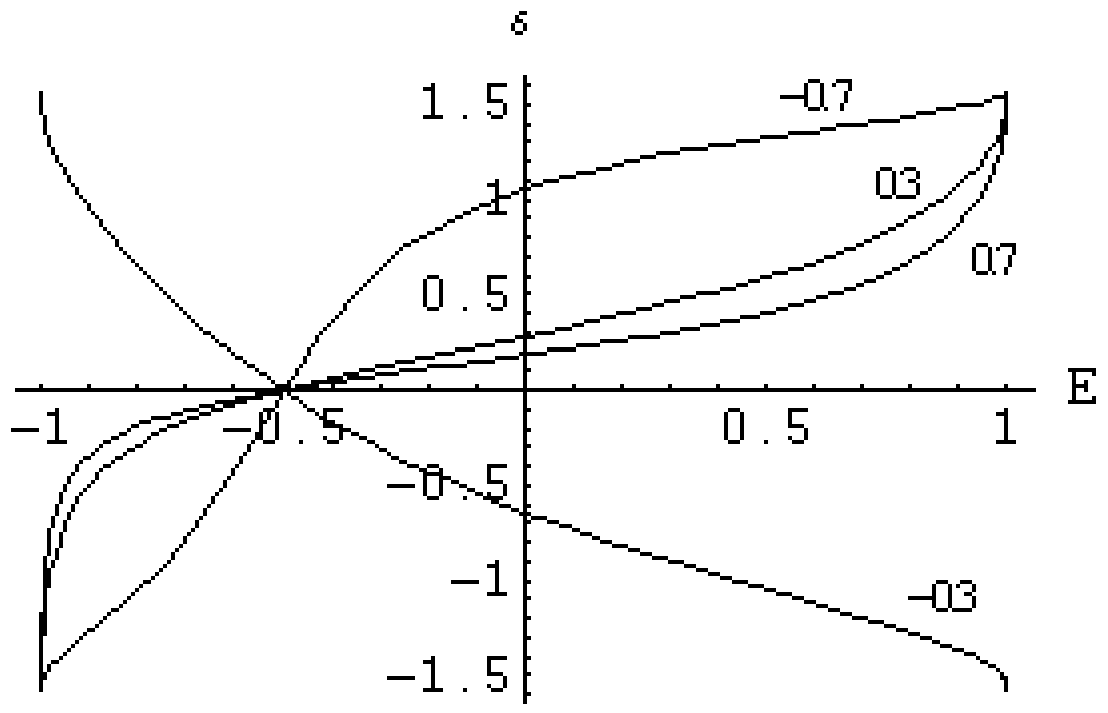

Fig.2

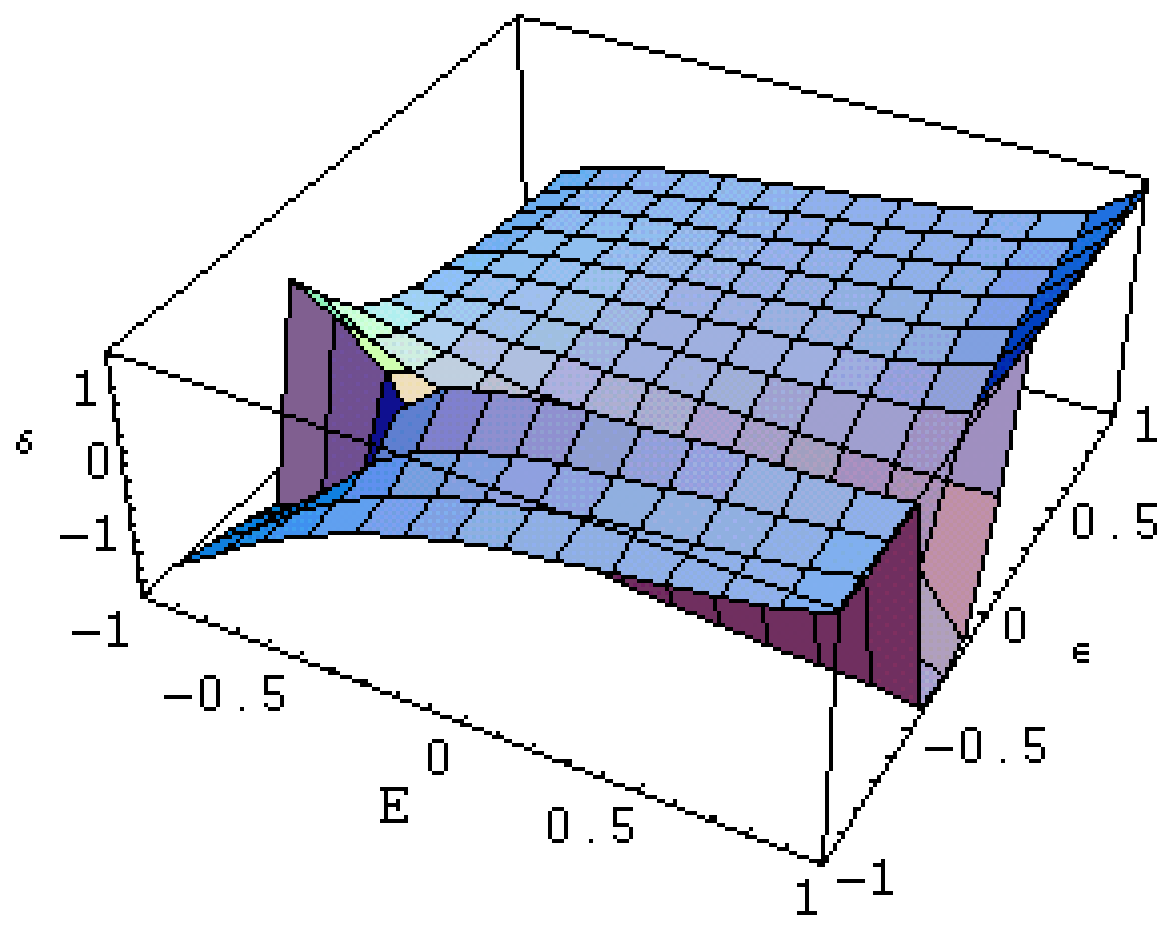

Fig.3

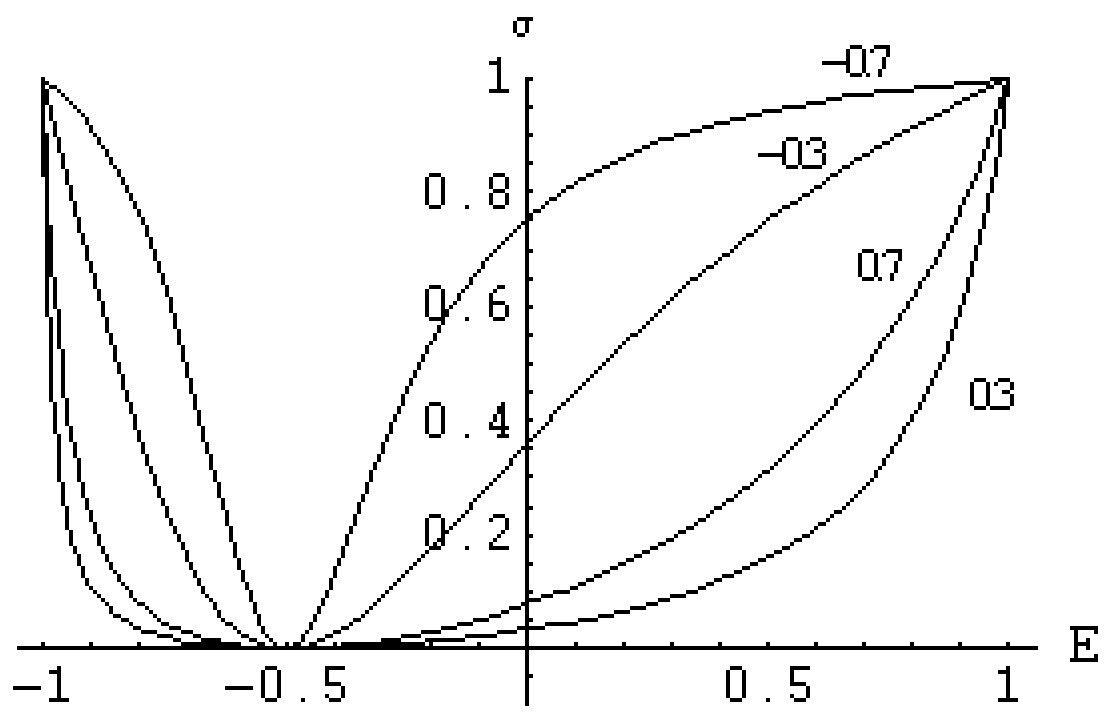

Fig.4

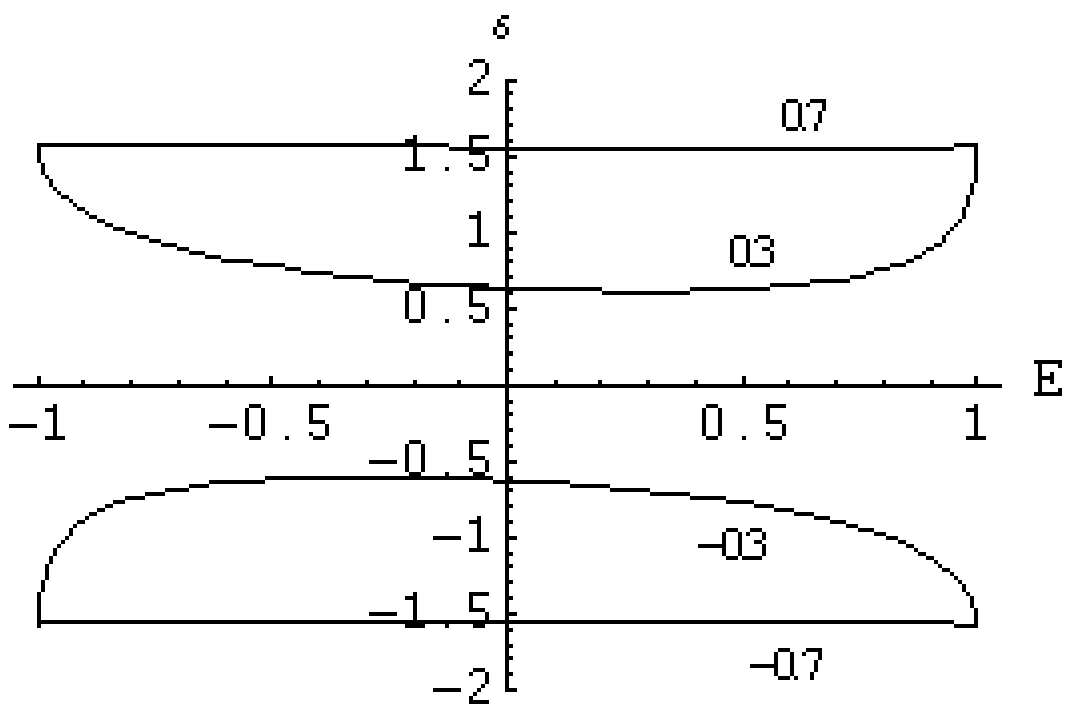

Fig.5

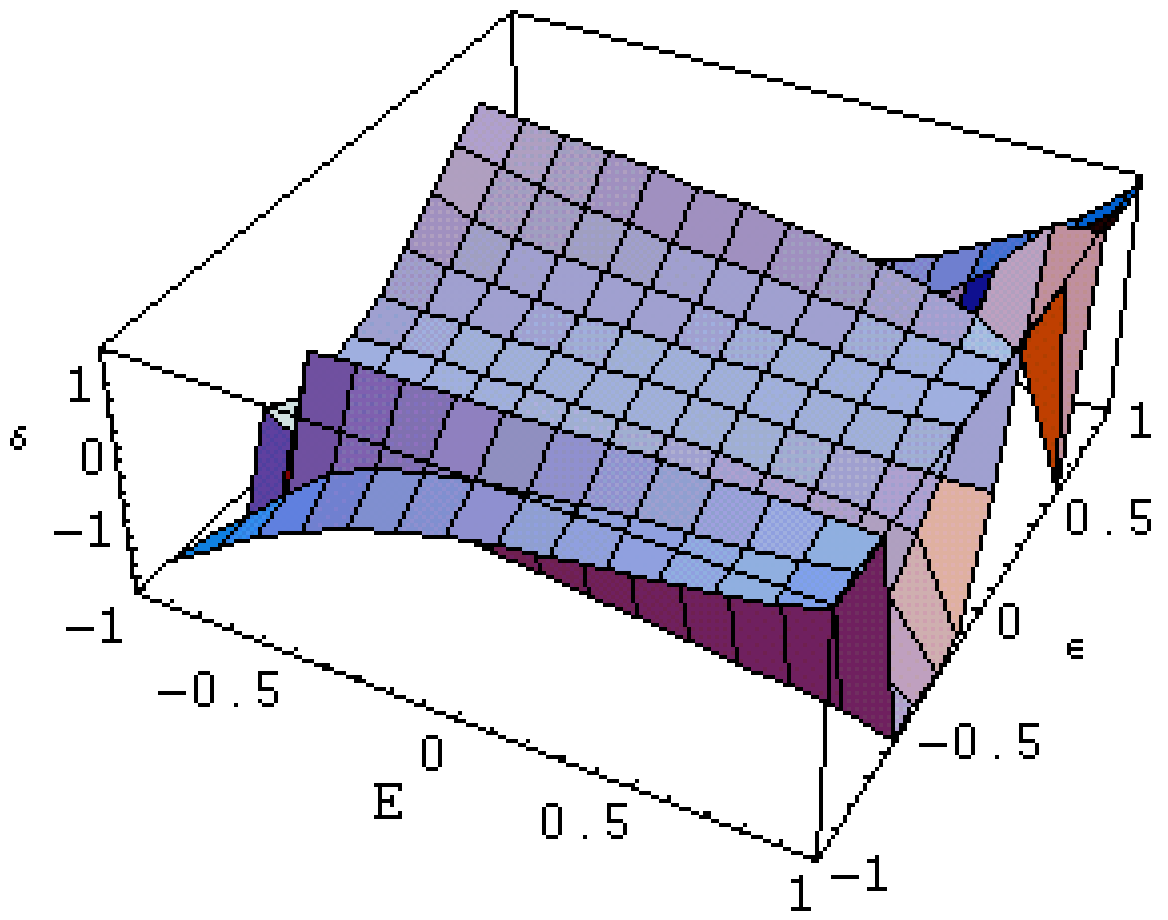

Fig.6

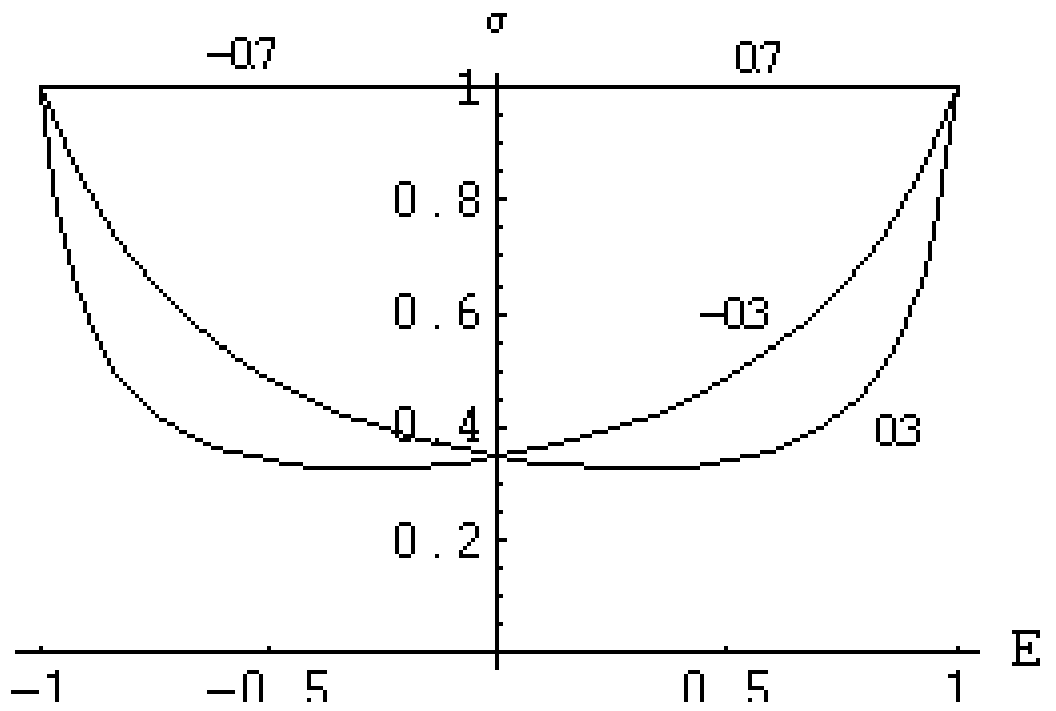

Fig.7

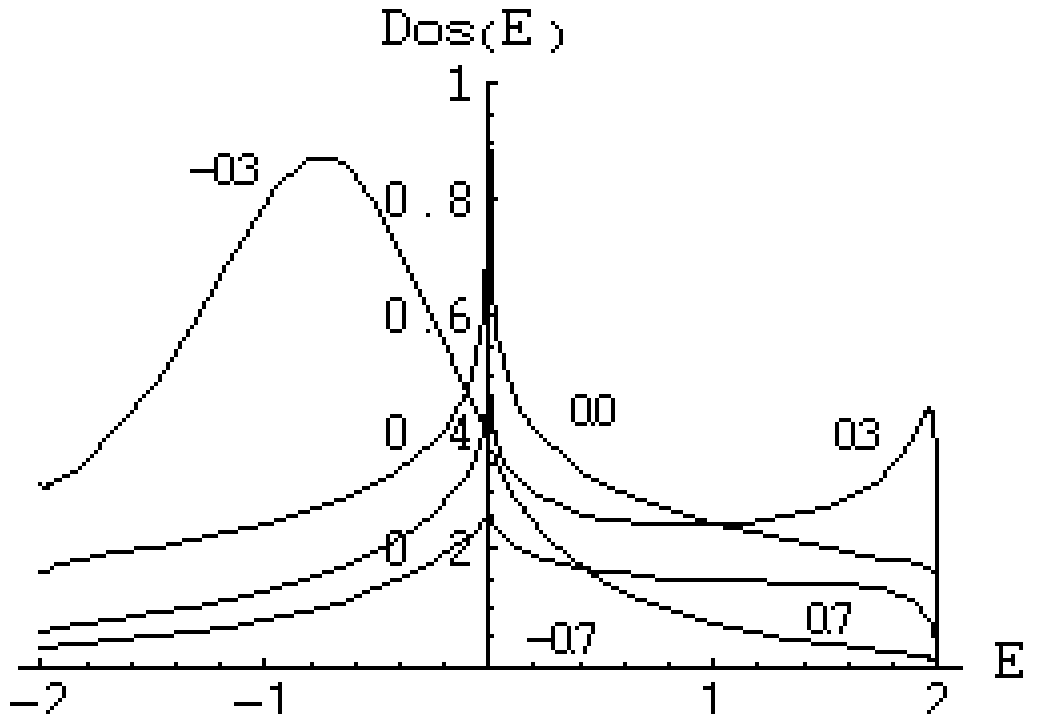

Fig.8

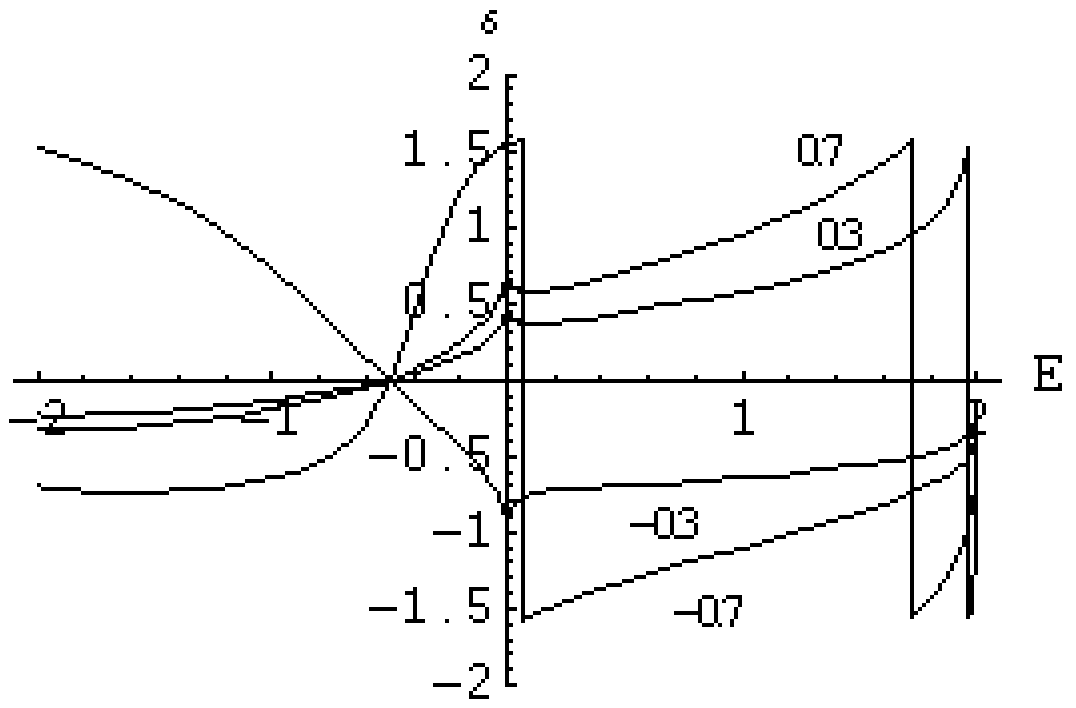

Fig.9

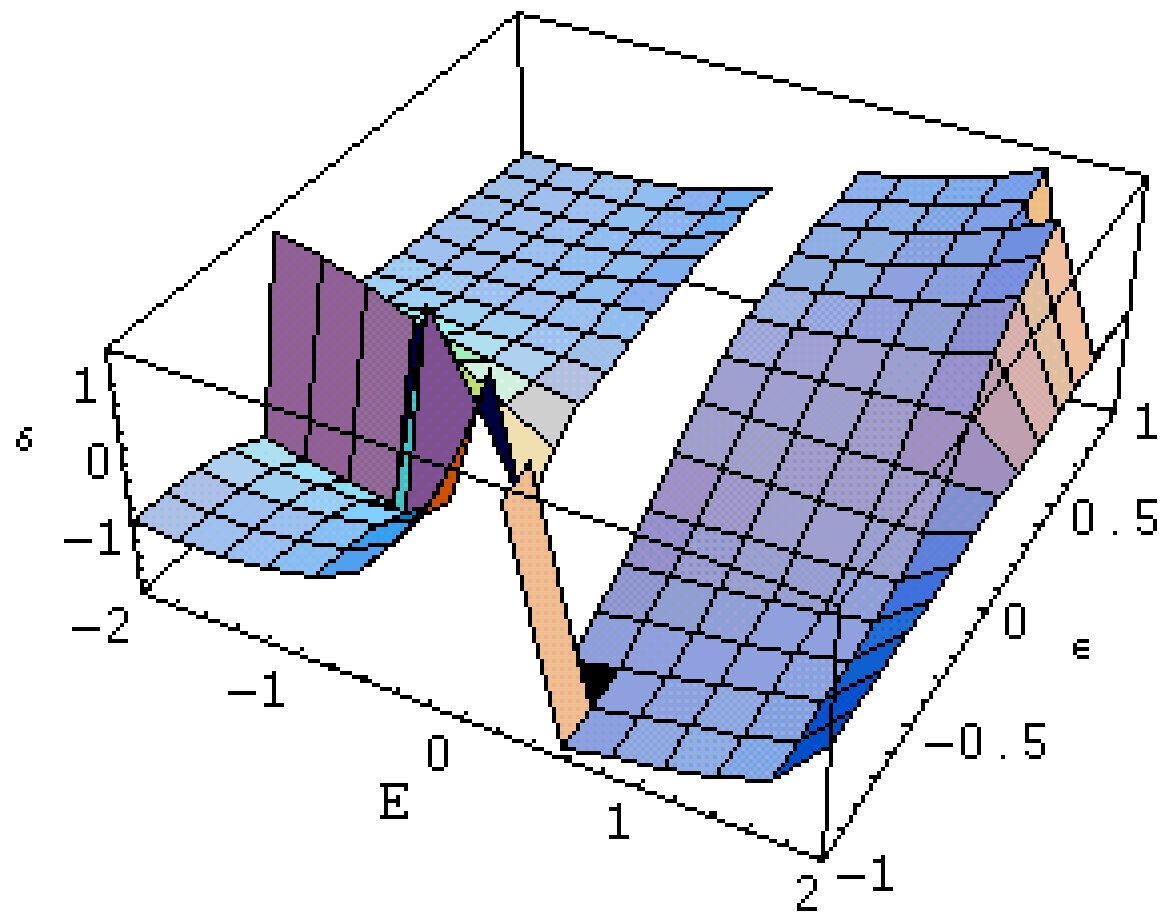

Fig.10

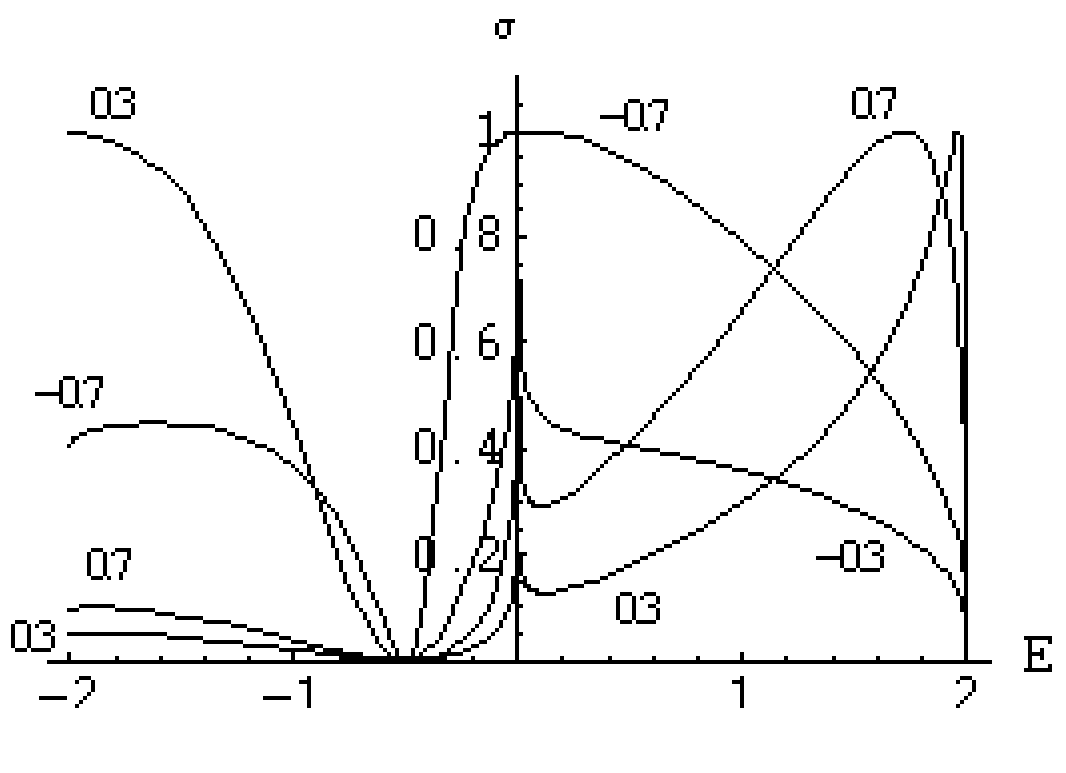

Fig.11

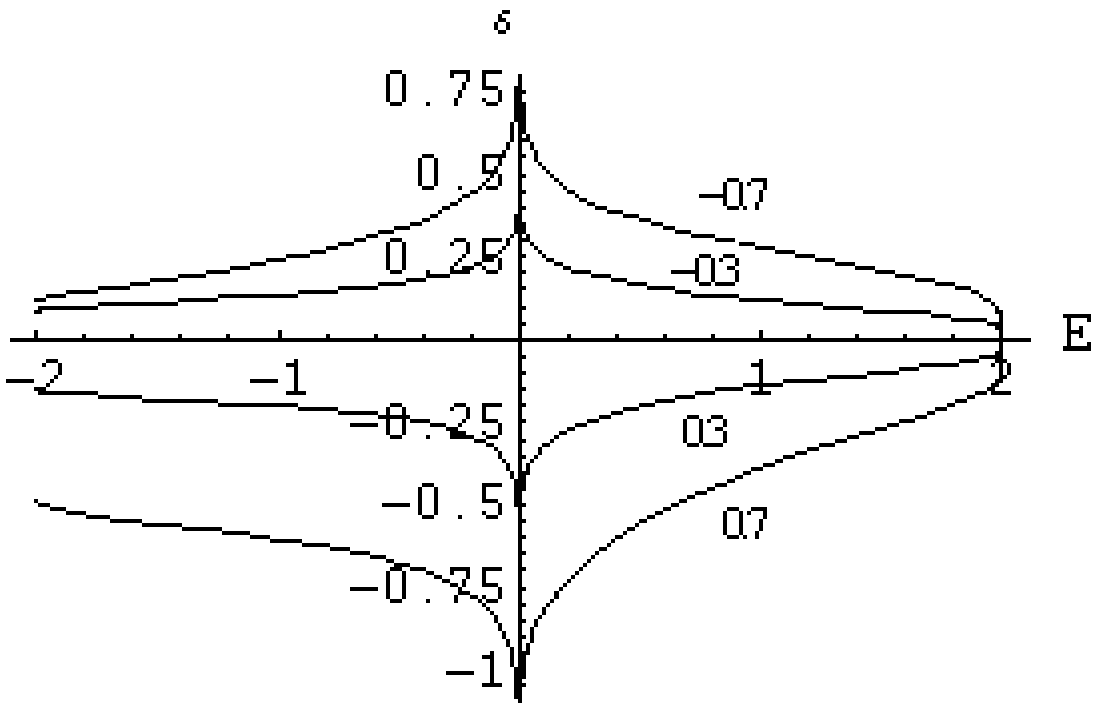

Fig.12

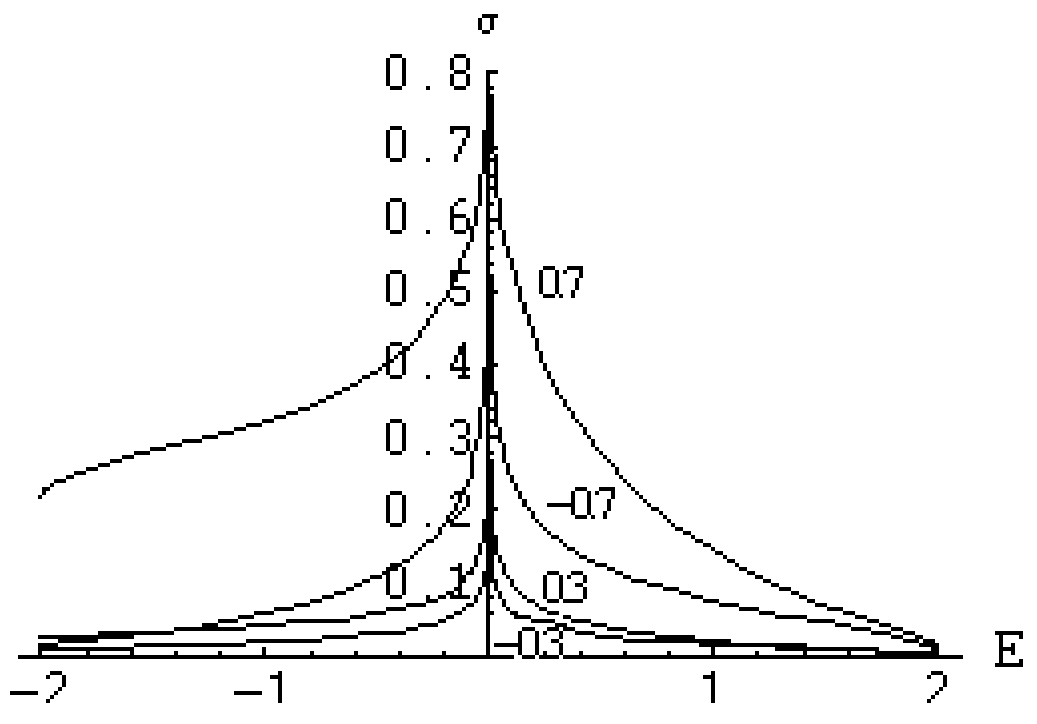

Fig.13

**References:**

1- E. N Economou, Green’s Functions in Quantum Physics (Springer – Verlag, Berlin 1983), second edition.

2- R. S. Hijjawi. 2002. Thesis (Ph.D). University of Jordan (unpublished).

3- R. S. Hijjawi, and J. M. Khalifeh. "Remarks on the lattice Green's Function, the General Glasser Case". J. Theo. Phys., **41**, 1769 (2002).

4- R. S. Hijjawii, J. H. Asad, A. Sakaji, and J. M. Khalifeh. " Lattice Green's Function for the face Centered cubic Lattice". Int. J. Theo. Phys., **(43)** 11:2299.

5- G. Rickayzen. Green's functions and Condensed Matter. Academic Press. London (1980).

6- B. M. mccoy, and T. T. wu. "two- Dimensional ising Model theory for $T \neq T_c$ : Green's function strings in n- point Function. Physical Review D, **(18)** 8: 1253 (1978).

7- E. Montroll, and G. weiss. " Random walks on Lattices". J. Math. Phys., **(4)** 1:241 (1965).

8- G. L. Montent. "Integrals Methods in the Calculation of Correlation Factors in Diffusion". Physical review B, **(7)** 3: 650 (1975).

9- N. W. Dalton, and D. W. Wood. " Critical Behavior of the Simple An Isotropic Heisenberg Model". Proceedings Physics Society (London), **(90)**: 4591 (1967).

10- G. F. Koster, and D. C. Slater. "Simplified Impurity Calculation". Physical Review, **(96)** 5: 1208 (1954).

11- V. Bryksin, and P. Kleinert. "Tight- Binding electron on a Disordered

Square Lattice in A magnetic Field". Zeits Physics B (Condensed Matter). **90**: 167 (1993).

12- J. Cserti. " Application of the Lattice Green's Function for Calculating The Resistance of Infinite Networks of Resistors". Am. J. Phys., **(68)**: 896 (2000).

13- J. Cserti, G. David, and A. Piroth. "Perturbation of Infinite Networks of Resistors". Am. J. Phys., **(70)**: 153 (2002).

14- J. H. Asad, R. S. Hijjawi, A. Sakaji, and J. M. Khalifeh. "Resistance Calculation for an Infinite Simple Cubic Lattice- Application of Green's Function". Int. J. Theo. Phys., **(43)** 11: 2223(2004).

15- J. H. Asad, R. S. Hijjawi, A. Sakaji, and J. M. Khalifeh. " Remarks on Perturbation of Infinite Networks of Identical Resistors ". Int. J. Theo. Phys., 44 (4): 481-494.

16- J. H. Asad. 2004. Thesis (Ph.D). University of Jordan (unpublished).

17- T. Horiguchi. " Lattice Green's Function for the Simple Cubic Lattice". J. Phys. Society Japan.**(30)** 5: 1261 (1971).

18- G. S. Joyce. "Lattice Green Function for the Simple Cubic Lattice". J. Phys. A: Math. Gen.**(5)**: L65 (1972).

19- M. Inoue. " Lattice Green's Function for the Body Centered Cubic Lattice". J. Math. Phys., **(16)** 4: 809 (1975).

20- T. Horiguchi, and T. Morita. " Note on the Lattice Green's Function for the Simple Cubic lattice". J. Phys. C, **(8)**: L232 (1975).

21- M. L. Glasser, and J. Boersma. " Exact values for the Cubic Lattice Green Functions". J. Phys. A: Math. Gen., **(33)** 28: 5017 (2000).

22- A. Sakaji, R. S. Hijjawi, N. Shawagfeh, and J. M. Khalifeh. " Remarks

on the Lattice Green's Function the Glasser Case". J. Math. Phys., **(43)**, 1 (2002).

23- A. Sakaji, R. S. Hijjawi, N. Shawagfeh, and J. M. Khalifeh. " Remarks on the Lattice Green's Function for the Body Centered Cubic Lattice". J. Theo. Phys., **(41)**, 973 (2002).

24- T. Morita. " Useful Procedure for Computing the Lattice Green's Function- Square, Tetragonal, and BCC Lattices". J. Math. Phys., (12), 1744 (1975).

25- S. Katsura, and S. Inawashiro. " Lattice Green's Functions for the Rectangular and the Square lattice at Arbitrary Points". J. Math. Phys., **(12)**, 1622 (1971).

26- T. Morita. " Use of a Recurrence Formula in computing the Lattice Green's Function". J. Phys. A: Math. Gen. **(8)**, 478 (1975).

27- I .S. Gradshteyn and I. M. Ryzhik, Tables of Integrals, Series, and Products (Academic, New York, 1965).

28- Bateman Manuscript Project, Higher Transcendental Functions, Vol. I,edited by A. Erdelyi *et al.*(McGraw-Hill, New York, 1963).

29- P. M. Morse and H. Feshbach, Methods of Theoretical Physics (McGraw-Hill, New York, 1953).